\documentstyle[12pt]{article}
\begin{document}
\begin{center} \Large Platelet Collapse Model of Pulsar Glitches\\
\vspace*{.15in}
P.D. Morley \\XonTech, Inc. \\6862 Hayvenhurst Ave.
\\ Van Nuys, CA 91406\\
\vspace*{.15in}
and \\
\vspace*{.15in}
Iv\'{a}n Schmidt\footnote{Work
supported in part by FONDECYT (Chile), contract 1931120} \\
Department of Physics \\ Universidad T\'{e}cnica Federico Santa Mar\'\i a \\
Casilla 110-V \\ Valpara\'\i so, Chile
\end{center}

\begin{abstract}
A platelet collapse model of neutron starquakes is introduced. It displays
self-organized criticality with a robust power-law behavior. The simulations
indicate a near-constant exponent, whenever scaling is present.
\end{abstract}
\normalsize

\vspace{.25in}

\begin{center}
PACS Numbers: 91.30.Px, 05.40.+j, 97.60Gb, 97.60Jd \end{center}
\vspace*{.2in}
\begin{center} Preprint USM-TH-64 \end{center}

\newpage

Bak, Tang and Wiesenfeld [1] introduced the concept of self-organized
criticality
(SOC): large dissipative dynamical systems tend to drive themselves to a
critical state without the need to fine-tune system parameters. SOC seems to
provide a natural explanation for the ubiquitous scale-free phenomena seen in
nature.
The critical state is a non-equilibrium state characterized by spatial
correlation functions exhibiting power-law behavior. Until the finite size
of the system comes into play, there is no intrinsic length scale in the
dynamics, giving rise to the observed scaling. Thus the outstanding feature
of SOC is the appearance of avalanches of dynamical activity of all sizes.
Recently Bak [2] has given examples of SOC in astrophysics, including light
emitted from quasars, emission of hard x-rays from solar flares and pulsar
glitches. In this paper we present a detailed model of pulsar glitches,
which realizes SOC,
characterized by partial or full collapse of the neutron star's surface
induced by crustal strain.

A pulsar glitch is an essentially instantaneous discontinuity of the
rotating neutron star's period to shorter values. Morley and Garc\'\i a-Pelayo
[3] have taken the world's pulsar glitch data and have shown
empirically that the probability $P(E)$ that a macro-glitch of energy $E$
can occur in a glitching pulsar has the scaling law
      \begin{equation} P(E) \sim E^{-\gamma} \end{equation}
where the critical exponent, $\gamma$, due to the small number of recorded
glitches,
is not a well-determined number, lying in the range $ +1< \gamma < +2$.

There are many hundreds of known pulsars, of which only a minute number
actually
glitch. Morley [4] has shown that a reasonable explanation of why pulsars
glitch is the following: some fraction of pulsars are embedded in nebula or in
regions
of high interstellar mass density which allows them to accrete matter. All
neutron stars and white dwarfs have an inherent instability induced by mass
accretion, due to their inability to hydrostatically change their
configuration.
As one of these stars accretes matter and goes to a higher mass, its radius
is also increasing which pushes the star to radial instability\footnote{Masses
of neutron stars and white dwarfs in stable equilibrium are monotonic
{\it decreasing} functions of their radius.}. This mass accretion stress is
added to the stress caused by the torque on the star, which causes the
pulsar to be an ellipsode of revolution for the wrong angular speed. Together,
these stresses induce tremendous strain in the neutron star's surface crust.
When the threshold level is exceeded, the crust relaxes, moving inward,
resulting
in a starquake. By conservation of angular momentum, this inward movement
of surface crust causes the pulsar to immediately speed-up, the pulsar glitch.

We now model this dynamical process. A pulsar has a misalignment of its
spin axis $\vec{S}$, with its magnetic dipole axis $\vec{M}$ (see fig.\ 1).
In the $\vec{M}$ frame of reference, Ruderman [5] has argued that the surface
of the star is divided into platelets which have varying surface sizes,
constrained by the requirement that the total magnetic flux threaded through
a platelet is approximately constant. We adopt the Olami et al. [6] cellular
automaton model of earthquakes to this starquake model. Each platelet
referenced
(i,j) is connected to neighbors by mean shear moduli $\mu_{s}$. The
displacement
of each platelet from its relaxed position is defined as $dr_{i,j}$. The total
radial stress on a given platelet is
      \begin{equation} S_{i,j} =\mu_{s}^{lo} [2dr_{i,j} -dr_{i-1,j}
-dr_{i+1,j}]
       + \mu_{s}^{la}[2dr_{i,j} -dr_{i,j-1}-dr_{i,j+1}]
       +\mu_{c}dr_{i,j} \; .  \end{equation}
In eq(2) $\mu_{s}^{lo}, \mu_{s}^{la}$ and $\mu_{c}$ are respectively the
longitudinal,
latitudinal shear moduli and the compression modulus. The latter stress
is associated with mass
accretion, while the former are shear stresses. The total stress on a platelet
increases uniformily (with a rate proportional to $ \frac{dR}{dt}$, where $R$
is the star's radius) until one platelet reaches its threshold value and
relaxes.
This triggers the starquake. The redistribution of stress after a platelet
collapse at position (i,j) is
      \begin{eqnarray} S_{i \pm 1,j} & \rightarrow & S_{i \pm 1,j} + \Delta
S_{i \pm 1,j} \nonumber \\
      S_{i,j \pm 1} & \rightarrow & S_{i,j \pm 1} + \Delta S_{i, j \pm 1} \\
      S_{i,j} & \rightarrow & 0 \nonumber \end{eqnarray}
where
     \begin{eqnarray} \Delta S_{i \pm 1,j} = \frac{\mu_{s}^{lo}}
     {2 \mu_{s}^{lo} + 2 \mu_{s}^{la} + \mu_{c}} S_{i,j} = \beta^{lo} S_{i,j}
\nonumber \\
     \Delta S_{i, j \pm 1}= \frac{\mu_{s}^{la}}{2 \mu_{s}^{lo} + 2 \mu_{s}^{la}
     + \mu_{c}} S_{i,j} = \beta^{la}S_{i,j} \; . \end{eqnarray}
For the purposes of this paper, we put $\beta^{lo}=\beta^{la}=\beta$.
Eqs(2)-(4) constitute the Olami et al.\ cellular automaton model [6]. We now
add the important changes intrinsic to starquakes. We designate $\lambda
(\theta_{i}, \phi_{j})$ as the surface platelet density function; that is, the
platelet whose center is at position (i,j) has an area $\lambda (\theta_{i},
\phi_{j})$ in the $\vec{M}$-system. Since the radial magnetic field is
$M_{R}=M_{0} \cos(\theta_{\vec{M}})$ we parametrize $\lambda(\theta_{i},
\phi_{j})$
as ($\theta_{i}$, $\phi_{j}$ are $\vec{M}$ system angles)
     \begin{equation} \lambda(\theta_{i},\phi_{j}) = \frac{C}{a +
\cos(\theta_{i})^{2}} \; .
     \end{equation}
The constant $C$ is determined by the area normalization condition. We adopt
a spherically shaped pulsar, and account for its eccentricity stresses
directly as seen below. To handle the boundary topology of a sphere, we
disconnect the pole and anti-pole points from their neighbors. By discretizing
the $\theta_{\vec{M}}$ variable into $M$ values, $C$ satisfies
$C \sum_{i=2}^{M-1} \frac{ \Delta x_{i}}{a + x_{i}^{2}} = 2(1-\Delta x /
(a +1))$, where $x=\cos(\theta_{\vec{M}})$, $\Delta x_{i}= \Delta x= 2/M$,
$x_{i} = \frac{2i-M-1}{M-1}$. $\phi$ is discretized into $N$ values by
$\Delta \phi = \frac{2 \pi}{N}$, so $\phi_{j}=j \Delta \phi$. There are thus
$N \times (M-2)$ platelets in the $\vec{M}$-system covering the star. In eq(5),
$a$ is the parameter which determines the ratio of the largest to smallest
platelet sizes
$\simeq \frac{1}{a + a^{2}}$. Since the eccentricities ($\epsilon,
\epsilon_{0}$)
of pulsars are minute,
we adopt a spherical configuration. To account for the centrifugal
strain of having $\epsilon$ deformation instead of $\epsilon_{0}$
(strain-free),
we introduce a platelet threshold strain, $S_{th}(\theta_{\vec{S}})$, which is
a function of $\sin(\theta_{\vec{S}})$, where $\theta_{\vec{S}}$ is the polar
angle in the $\vec{S}$-system:
      \begin{equation} S_{th}(\theta_{\vec{S}})_{i,j} = S_{T}(1- \alpha
      |\sin(\theta_{\vec{S}})|_{i,j}) \; . \end{equation}
$\alpha$ in eq(6) is the parameter which accounts for the reduced threshold for
platelet collapse induced by the centrifugal strain. All stresses are in
units of $S_{T}$, which can be put equal to 1. By geometry
$|\sin(\theta_{\vec{S}})|_{i,j} = (1-(\sin(\psi) \cos(\phi_{j})
\sqrt{1-x_{i}^{2}}
+ x_{i} \cos(\psi))^{2})^{1/2}$, where $\psi$ is the angle between $\vec{S}$
and $\vec{M}$. The last major change is the new boundary conditions. All
platelets except those ringing the pole and anti-pole have continuous boundary
conditions
of four neighbors. Those ringing the poles have only three. The parameters
of the model are $\psi, a, \alpha, \beta$ and the number $N \times (M-2)$. The
simulation proceeds as follows:
\begin{enumerate}
\item[1)] Initalize all platelet stresses to a random value between 0 and
$S_{th}(\theta_{\vec{S}})_{i,j}$.
\item[2)] If any $S_{i,j} \geq S_{th}(\theta_{\vec{S}})_{i,j}$, the platelet
relaxes. Redistribute the stress on $S_{i,j}$ according to eq(3) and add
the areas of the collapsed platelets.
\item[3)] Repeat step 2) until the starquake is fully evolved.
\item[4)] Locate the platelet with the largest strain such that
$\Delta S = S_{th}(\theta_{\vec{S}})_{i,j} - S_{i,j}$ is smallest.
\item[5)] Add $\Delta S$ to all platelets (uniform mass accretion) and return
to step 2).
\end{enumerate}
The energy of a starquake comes from gravity. We assume the platelets fall a
fixed
relative distance $\frac{\Delta R}{R}$, so the energy is just proportional
to the area of the collapsed platelets: $E \sim \sum_{collapsed \; i,j} \lambda
(\theta_{i},\phi_{j})$. Interestingly enough, there are rare starquakes where
more than the total surface area collapses due to the near-threshold level of
all platelets and the continuous boundary conditions.

In fig.\ 2, we present $\log \frac{dN}{dE}$ versus $\log E$,
with the listed parameters\footnote{All
runs had 999999 starquakes.}. No SOC is present. This can be
understood rather simply: the value of $a$ means there is a disproportionate
area size distribution among platelets which means that the starquakes are
dominated by the large platelet collapses. By changing $a$ to a value where
the area density is not greatly varying, fig.\ 3, SOC is realized.
Simulations indicate that SOC is present for all small $a$ values and the
exponent is almost independent of the values of $\psi$ and $\alpha$. Only
variation in $\beta$ produces a weak modification of the scaling exponent,
fig. 4.

Our computer results strongly support SOC in the phenomena of pulsar glitches.
They indicate that mass accretion is the critical element driving the neutron
star instability. The critical exponent should lie between $1.5 < \gamma <
2$, depending only on the ratios of shear and compression moduli.

\newpage

\begin{center} {\bf Figure Captions} \end{center}

\begin{enumerate}
\item[1.] Platelets on the rotating neutron star with misaligned spin $\vec{S}$
and magnetic dipole $\vec{M}$ axis.
\item[2.] $\log \frac{dN}{dE}$ versus $\log E$, for $\psi$=0.5, $a$=0.01,
$\alpha$=.1,
N = M = 50 and $\beta$=0.2.
\item[3.] Same as fig.\ 2, except $a$=2.0.
\item[4.] Critical exponent $\gamma$ versus stress parameter $\beta$.
\end{enumerate}
\end{document}